\documentclass[preprint,showpacs,preprintnumbers,amsmath,amssymb]{revtex4}
\usepackage{graphicx}
\usepackage{dcolumn}
\usepackage{bm}

\begin{document}            


\title{Optimizing single-photon-source heralding efficiency  at 1550 nm using periodically poled lithium niobate}

\author{S. Castelletto, I.P. Degiovanni,\\ V. Schettini, and A. Migdall$^{*}$}


\address{IEN G. Ferraris, Photometry Dept, Turin 10134, Italy,
$^{*}$NIST, Optical Technology Division, Gaithersburg, MD USA
20899-8441}



\begin{abstract}
 We explore the feasibility of using
high conversion-efficiency periodically-poled crystals to produce
photon pairs for photon-counting detector calibrations at 1550 nm.
The goal is the development of an appropriate parametric
down-conversion (PDC) source at telecom wavelengths meeting the
requirements of high-efficiency pair production and collection in
single spectral and spatial modes (single-mode fibers). We propose a
protocol to optimize the photon collection, noise levels and the
uncertainty evaluation. This study ties together the results of our
efforts to model the single-mode heralding efficiency of a
two-photon PDC source and to estimate the heralding uncertainty of
such a source.
\end{abstract}

\maketitle

\section{Introduction}
Parametric down-conversion (PDC) consumes pump photons and
produces light with a two-photon field description
\cite{klyshkobook}. This two-photon light, which allows one photon
to indicate or herald the existence of its twin, has a key role in
applications such as quantum radiometry \cite{MCD02}, where it
makes possible an independent primary standard method for direct
photon-counting detector calibrations. In addition, this
two-photon light is key to improve single-photon source schemes
that have significant advantages over the attenuated laser sources
often employed in quantum information \cite{TBZ00,KLM01}
applications.
 To realize a  single-photon source
\cite{fasel04} appropriate for these applications, we consider two
key issues- a) efficient photon pair production and b) collection of
the heralded photon in a single-mode fiber, which is of interest for
telecom applications. Periodically poled crystals (PPLs) have been
used for efficient pair production \cite{MAD04}, but have not yet
been used for metrology. While metrology applications would
certainly benefit from the high pump-conversion efficiency of PPLs,
characteristics such as the level of background fluorescence may
limit the uncertainty that can be achieved.

As for the collection efficiency of PDC light into a single-mode
fiber, the goal has been to understand how the detection of one
photon of a PDC pair in a well defined spatial and spectral mode,
defines its partner's spatial and spectral mode that can then be
efficiently collected. Recent efforts have included
 experimental and theoretical work in both continuous wave (CW)
\cite{kurtsiefer01,CDSWM04,DLT05} and pulse-pumped
\cite{bovino03,PJF04,CDS05} bulk crystal PDC, and have highlighted
the critical problem of coupling the PDC photons into single-mode
fibers (SMF) due to the difficulty of spatial and spectral matching
of the photon pairs. The results show that optimization of the
heralding efficiency is very sensitive to spectral and spatial mode
selection.

Here we describe a PDC source and a measurement technique and
compare it to other configurations. We present theoretical and
experimental characterizations of the photon collection problem and
suggest possible solutions.
\section{Experimental setup}
\begin{figure}[tbp]
\begin{center}
\includegraphics[angle=0,width=15 cm]{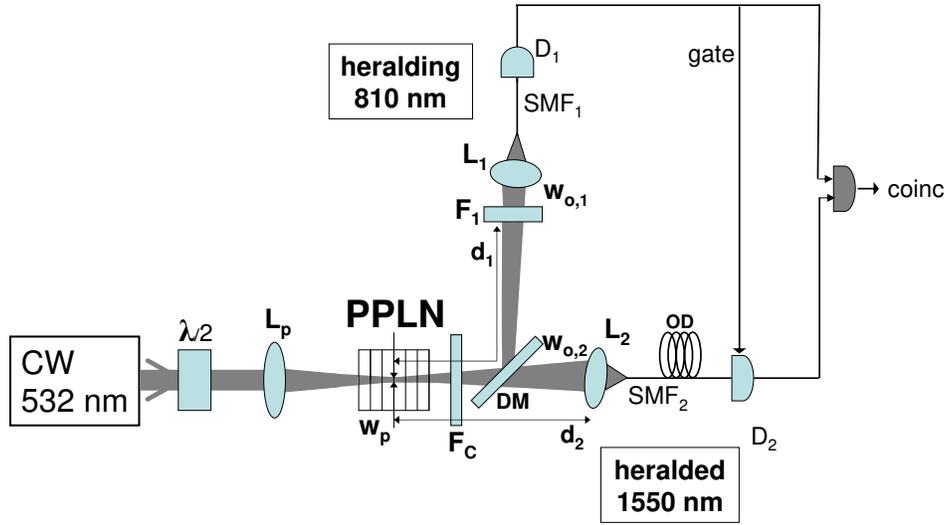}
\end{center}
\caption{Setup to herald single-photons from CW PDC from a PPLN
crystal. } \label{fig:Fig1Met}
\end{figure}

We pumped (Fig. \ref{fig:Fig1Met}) a 5 mm long periodically poled
MgO-doped lithium niobate (PPLN) crystal with a CW laser at 532
nm. We used noncritical phase-matching (90$^{\circ}$
phase-matching angle) with a 7.36 $\mu$m poling period to produce
810 nm and 1550 nm photon pairs at external angles of 1$^\circ$
and 2$^\circ$. We achieved fine tuning by adjusting the crystal
temperature near 131$^{\circ}$ C. Lens $\text{L}_{\text{p}}$ in
the pump beam produced a gaussian beam waist of
$w_{\text{p}}\simeq$144$\mu$m at the crystal; cutoff filter
F$_{\text{C}}$ blocked the pump laser; and dichroic mirror DM
separated the 810 nm (beam 1) and 1550 nm (beam 2) photons. An
extra interference filter ($\text{F}_{1}$) at 810 nm with a full
width half-maximum (FWHM) of 10 nm suppressed extra fluorescence
from the PPLN to reduce background heralding counts, while the SMF
geometry imposed spectral selection with a heralding bandwidth,
${\Delta}_{1}$ of $\approx$2 nm FWHM (see Table I). L$_{1,2}$ were
aspheric coupling lenses of focal length 8 mm, AR-coated for 810
nm and 1550 nm. A monochromator in the heralding arm measured its
spectral width. The heralding arm was routed to a SMF and then to
D$_{1}$, a Si Avalanche Photodiode (APD), while the heralded arm
also coupled to a SMF, was sent to D$_{2}$, an InGaAs APD
operating in gated mode. We gated the InGaAs detector with the
detection of an 810 nm photon in the heralding arm and adjusted
the delay of the heralded arm with appropriate length SMF and
electronic cables. Calibrating the InGaAs detector, including the
SMF and coupling lens optical losses against a conventional
detector standard using an attenuated laser source, yielded a
detection efficiency of $\eta_{\text{det}}= 9.8\% \pm 0.5\%$
(\emph{k}=1, absolute uncertainty).
\section{Measurement technique}
The heralded D$_{2}$ detector efficiency as measured by the
 PDC source is given by
 \begin{equation}\label{eq.1}
\eta_{\text{det}}=\chi_{\text{D}}/ (\chi_{\text{P}} \cdot
\tau_{\text{opt}}\cdot \tau_{\text{SMF-lens}})
\end{equation}
where $\chi_{\text{P}}$ is the heralding efficiency (single-mode
preparation efficiency, discussed in the next section)
\cite{CDSWM04,CDSM05}, $\tau_{\text{opt}}$ is the overall heralded
arm optical transmittance
 (including PPLN, F$_{\text{C}}$, and DM), and $\tau_{\text{SMF-lens}}$ is the optical transmittance of L$_{2}$ and the SMF on the heralded
 arm.
$\chi_{\text{D}}$ is the raw D$_{2}$ channel detection efficiency,
directly
 measured according to
  \begin{equation}\label{eq.3}
\chi_{\text{D}}=
\frac{P_{\text{coinc}}-P_{\text{uncorr}}}{(1-P_{\text{uncorr}})(1-P_{\text{backgnd}}^{\text{heralding}})},
\end{equation}
where $P_{\text{coinc}}$ is the probability of coincidence counts
per gate, $P_{\text{uncorr}}$ is the probability of uncorrelated
or accidental coincidence counts per gate (determined by changing
the heralding delay so the detection gate misses heralded
photons), and $P_{\text{backgnd}}^{\text{heralding}}$
 is the probability of gating counts produced by uncorrelated photons and
 dark counts in the heralding arm.
\subsection{Raw detection efficiency}
To understand Eq.(\ref{eq.3}) we note that when D$_{1}$ fires,
D$_{2}$ is gated for a duration $T$. We define the probability of
D$_{2}$ not firing during $T$ as $p^{0}_{T}$. We assume that the
``events" that make D$_{2}$ fire have a poissonian distribution
(ie., uniform in time). D$_{1}$ can fire for either background or
PDC photons, and the associated probabilities are
$P^{\text{heralding}}_{\text{backgnd}}$ and
$P^{\text{heralding}}_{\text{PDC}}=1-P^{\text{heralding}}_{\text{backgnd}}$.
The probability that D$_{2}$ fires for true correlated photons is
$P_{\text{coinc}}^{\text{heralded}}=
P^{\text{heralding}}_{\text{PDC}} p^{0}_{T/2} \chi_{\text{D}}$,
where $p^{0}_{T/2}$ is the probability for the heralded detector
to fire for the correlated photon arriving at \emph{T}/2, as
opposed to firing during the first half of the gating time. (We
assume that correlated photons arrive exactly at \emph{T}/2.) The
probability that the heralded detector fires for accidental events
is
\begin{equation}
P_{\text{uncorr}}=(1-p^{0}_{T})P^{\text{heralding}}_{\text{backgnd}}+
(1-p^{0}_{T/2})P^{\text{heralding}}_{\text{PDC}} +
p^{0}_{T/2}(1-\chi_{\text{D}})(1-p^{0}_{T/2})P^{\text{heralding}}_{\text{PDC}}.
\end{equation}
The first term is the probability of an accidental coincidence due
to D$_{1}$ firing for a background event and D$_{2}$ firing for
any event. The second and third terms are for D$_{1}$ firing for a
PDC event and D$_{2}$ firing due to a background event either
before or after the the arrival of the PDC photon at \emph{T}/2.
For D$_{2}$ to fire after \emph{T}/2 it cannot have fired due to a
background photon before \emph{T}/2 or a PDC photon at \emph{T}/2.
Those factors make up the third term. The total coincidence
probability is
$P_{\text{coinc}}=P_{\text{coinc}}^{\text{heralded}}+P_{\text{uncorr}}$.
Considering that for poisson statistics
$p^{0}_{T}=p^{0}_{T/2}\times p^{0}_{T/2}$, and inverting the above
equations, we obtain Eq.(\ref{eq.3}), where
$P_{\text{uncorr}}=1-p^{0}_{T}$.
\subsection{Estimate and uncertainty}
We develop an estimate of the raw detection efficiency and the
associated uncertainty following Ref.\cite{CDR02}. The terms
$P_{\text{coinc}}$, $P_{\text{uncorr}}$, and
$P_{\text{backgnd}}^{\text{heralding}}$ in Eq.({2}) are all
independent statistical variables, therefore the estimate of the
raw detection efficiency is
\begin{equation}
\langle \chi _{\text{D}}\rangle=\left \langle \label{eq.4}
\frac{1}{1-P_{\text{backgnd}}^{\text{heralding}}}\right\rangle
\times \left[1-\left \langle\frac{1}{1-P_{\text{uncorr}}}\right
\rangle(1-\langle P_{\text{coinc}}\rangle)\right].
\end{equation}
Applying the maximum likelihood model estimator to the random
variables $P_{\text{coinc}}$ and $P_{\text{uncorr}}$, the
probability of $M_{\text{coinc}}$ coincidence counts given
$M_{\text{heralding}}$ heralding counts is
\begin{equation}
P(M_{\text{coinc}}|M_{\text{heralding}},
p)=\frac{M_{\text{heralding}}!}{M_{\text{coinc}}!
(M_{\text{heralding}}-M_{\text{coinc}})!}\times
p^{M_{\text{coinc}}}
(1-p)^{M_{\text{heralding}}-M_{\text{coinc}}},
\end{equation}
where $p$ is the parameter to estimate. The maximum likelihood
function is then
\begin{equation}\label{ML} \nonumber
L(M_{\text{coinc}}, M_{\text{heralding}}|
p)=P(M_{\text{coinc}}|M_{\text{heralding}}, p)\times
 P(M_{\text{heralding}}),
\end{equation}
where $P(M_{\text{heralding}})$ is the distribution probability of
heralding counts, which is not known. The minimum of the function
$L(M_{\text{coinc}}|M_{\text{heralding}}, p)$ gives the estimate
of $P_{\text{coinc}}$, without the need of
$P(M_{\text{heralding}})$. It can be shown that $\langle
P_{\text{coinc}}\rangle=\langle
M_{\text{coinc}}/M_{\text{heralding}}\rangle$. Less trivial is the
estimate of $\langle P_{\text{uncorr}}\rangle$. We must estimate
$1/(1-P_{\text{uncorr}})$, as is clear from Eq.(\ref{eq.4}), so we
introduce the parameter $p'$ as $p=(p'-1)/p'$. The maximum
likelihood function is in this case
\begin{equation}\label{ML2}
L(M_{\text{uncorr}},M_{\text{heralding}}^{\text{delayed}}| p')=
P(M_{\text{uncorr}}|M_{\text{heralding}}^{\text{delayed}}, p)
P(M_{\text{heralding}}^{\text{delayed}}),
\end{equation}
which we minimize to obtain
$\langle(1/(1-P_{\text{uncorr}})\rangle=\langle
M_{\text{heralding}}^{\text{delayed}}/(M_{\text{heralding}}^{\text{delayed}}-M_{\text{uncorr}})\rangle$.
We distinguished $M_{\text{heralding}}$ from
$M_{\text{heralding}}^{\text{delayed}}$ because we measure
heralding counts in two configurations, one with the presence of
true coincidence photons and one in absence of coincidence when an
electronic delay line is applied to the heralding arm.
$M_{\text{uncorr}}$ are the coincidences corresponding to
uncorrelated events. For $P_{\text{backgnd}}^{\text{heralding}}$
the maximum likelihood model is not applicable because we cannot
measure it at the same time as the background counts
$M_{\text{backgnd}}$ and $M_{\text{heralding}}$, therefore we
apply $\langle P_{\text{backgnd}}^{\text{heralding}}
\rangle\approx\langle M_{\text{backgnd}}\rangle/\langle
M_{\text{heralding}}\rangle$. This implies that $\langle
P_{\text{backgnd}}^{\text{heralding}} \rangle$ must be close to 0,
to have no impact on the estimate of the raw detection efficiency.
We finally estimate
\begin{equation} \langle \chi_{\text{D}}\rangle=
\label{eq.5} \frac{1}{1-\frac{\langle
M_{\text{backgnd}}\rangle}{\langle M_{\text{heralding}}\rangle}}
\times \left[1-\left\langle\frac{
M_{\text{heralding}}^{\text{delayed}}}{M_{\text{heralding}}^{\text{delayed}}-M_{\text{uncorr}}}\right
\rangle \left(1-\left \langle
\frac{M_{\text{coinc}}}{M_{\text{heralding}}} \right
\rangle\right)\right],
\end{equation} with the uncertainty given by gaussian uncertainty
propagation for each random variable considered independently.

\subsection{Heralding efficiency}
The heralding efficiency $\chi_{\text{P}}$ is the efficiency of
preparing a photon in the heralded channel in a definite spectral
and spatial mode, by specific mode selection of the heralding arm.
It quantifies how well the collection system geometrically catches
photons correlated to those seen by D$_{1}$. To calibrate a
SMF-coupled detector, the heralding efficiency must be optimized
and estimated. It has been estimated in ref. \cite{CDSWM04} for a
bulk crystal. For PPLN \cite{CDSM05} it is
\begin{equation}\label{eq.2}
\chi_{\text{P}}=\frac{4~ w_{\text{p}}^{2} w_{\text{o,1}}^{2}
w_{\text{o},2}^{2} (w_{\text{o},1}^{2}+w_{\text{p}}^{2})
}{(w_{\text{o},2}^{2} w_{\text{p}}^{2}+
w_{\text{o},1}^{2}~(w_{\text{o},2}^{2}+
w_{\text{p}}^{2}))^{2}}\times
\frac{\Delta_{2}}{(\Delta_{1}^{2}+\Delta_{2}^{2})^{^{\frac{1}{2}}}}
\frac{f(c_{1},c_{2})}{f(s_{1},s_{2})} ,
\end{equation}
where $w_{\text{o},1,2}$ are the mode waists of the fibers at the
crystal, and $\Delta_{1,2}$ are the FWHM of the spectral
distribution selected geometrically by the SMFs.
 $f$ is a correction function accounting for crystal
length ($L$) and PDC crystal frequency dispersion given by
\cite{CDS05}
\begin{equation}
f(p,q)=\frac{\int^{1}_{0} \mathrm{d}x~ e^{-p
x^{2}+\frac{q^{2}x^{2}}{4 p}}(\mathrm{Erf}[\frac{q x}{2
\sqrt{p}}]-\mathrm{Erf}[\frac{-2 p+q x }{2 \sqrt{p}}])} {\sqrt{p}}.
\end{equation}
The parameters for this case are:
\begin{eqnarray}\label{coefficients}
c_{1}&=&c_{2} + \frac{ L^{2}(
w_{\text{p}}^{2}\alpha_{\text{s}}^{2}+ w_{\text{o},2}^{2}
\tan\theta_{\text{i}}^{2}+
w_{\text{o},1}^{2}(\alpha_{\text{s}}+\tan\theta_{\text{i}})^{2})}{
[w_{\text{o},2}^{2} w_{\text{p}}^{2}+
w_{\text{o},1}^{2}(w_{\text{p}}^{2}+w_{\text{o},2}^{2})]};
\\ \nonumber
 c_{2}&=& \frac{L^{2} D_{\text{is}}^{2}}{ a^{2}}
\frac{(\Delta_{1}^{2}
\Delta_{2}^{2})}{(\Delta_{1}^{2}+\Delta_{2}^{2})};\\  \nonumber
s_{1}&=& \frac{L^{2} D_{is}^{2} \Delta_{1}^{2}}{ a^{2}} +
  \frac{
L^{2}( w_{\text{p}}^{4}\alpha_{\text{s}}^{2}+
w_{\text{o},1}^{4}(\alpha_{\text{s}}+
\tan\theta_{\text{i}})^{2})}{ 2
w_{\text{o},1}^{2}w_{\text{p}}^{2}(w_{\text{p}}^{2}+w_{\text{o},1})}+
  \frac{
L^{2}( 2
w_{\text{o},1}^{2}w_{\text{p}}^{2}(\alpha_{\text{s}}^{2}+\alpha_{\text{s}}\tan\theta_{\text{i}}+\tan\theta_{\text{i}}^{2}))}{
2 w_{\text{o},1}^{2}w_{\text{p}}^{2}(w_{\text{p}}^{2}+w_{\text{o},1}^{2})};\\
\nonumber s_{2}&=& \frac{2 L^{2} D_{\text{is}}^{2}}{ a^{2}} +
  \frac{
L^{2}( w_{\text{p}}^{2}, \alpha_{\text{s}}+
w_{\text{o},1}^{2}(\alpha_{\text{s}}+
\tan\theta_{\text{i}})^{2})}{
w_{\text{o},1}^{2}w_{\text{p}}^{2}(w_{\text{p}}^{2}+w_{\text{o},1})}.
\end{eqnarray}
Here, $a=2 \sqrt{ \ln(2)}$ converts between the FWHM and the
gaussian profile 1/e$^2$ radius. The other terms are
$D_{\text{is}}
=D_{\text{i}}(\cos\theta_{\text{i}}-\sin\theta_{\text{i}}
\tan\theta_{\text{i}}) -
D_{\text{s}}(\cos\theta_{\text{s}}+\sin\theta_{\text{s}}\tan\theta_{\text{s}})$,
$D_{\text{pi}} = -D_{\text{i}}
(\cos\theta_{\text{i}}+\sin\theta_{\text{i}}
\tan\theta_{\text{i}}) + D_{\text{p}}$, with
$D_{\text{i,s}}=\frac{\textrm{d}
n_\mathrm{\text{i,s}}(\omega_\mathrm{i},\phi_{\mathrm{o}})\omega_\mathrm{i,s}/c}{\textrm{d}\omega_\mathrm{i,s}}
       |_{\Omega_\mathrm{i,s}}$, $D_{\text{p}}= \frac{\textrm{d} n_\mathrm{p}(\omega_\mathrm{p},\phi_{\mathrm{o}}
)\omega_\mathrm{p}/c}{\textrm{d}\omega_\mathrm{p}}
           |_{\Omega_\mathrm{p}}$. $\alpha_{\text{s}}=-\cos\theta_{\text{s}}
           \tan\theta_{\text{i}}+\sin\theta_{\text{s}}$. $\theta_{\text{i,s}}$ are the associated idler and signal
           emission angles
           around $\phi_{\text{o}}=\pi/2$ (the phase-matching angle in a non-critical
phase-matching configuration),
            and $n_\mathrm{i,s,p}(\omega_\mathrm{i,s,p},\phi_{\text{o}})$ are the
indices of refraction (all are e-rays) at the three frequencies.
$\Omega_{\text{i,s,p}}$ are the central angular frequencies.

For our experiment the fiber mode field diameters are
MFD$_{1}$=3.9 $\mu$m for the heralding arm and MFD$_{2}$=5.6
$\mu$m for the heralded arm, giving mode waists
$w_{\text{o},1,2}=M_{1,2}$ MFD$_{1,2}$ at the crystal according to
the magnification ($M_{1,2}$) used. Because of this, we identify a
SMF as a spectral filter with a gaussian spectral distribution
given by
$\widetilde{I}_{\text{s,i}}(\omega_{\text{s,i}}-\Omega_{\text{s,i}})\propto
e^{(-\frac{a^2}{\Delta_{1,2}^2}
(\omega_{\text{s,i}}-\Omega_{\text{s,i}})^{2})}$. The component of
spectral width due to geometric selection, $\Delta_{1,2}$ is given
by the FWHM angular collection $\Delta \theta_{1,2}=
a\frac{\lambda_{\text{i,s}}}{ \pi w_{\text{o},1,2}}$  and the
spectral/angular spread of the PDC,
$\theta_{\text{i,s}}(\lambda_{\text{s,i}})$ around the central
wavelength $\lambda_{\text{i,s}}$. Here, because the
non-degenerate PPLN configuration and the bandwidth estimation are
critical, we measured it with a monochromator in the heralding
arm. We spectrally scanned the heralding single counts at three
different values of $w_{\text{o},1}$ (Table I). We also
theoretically estimated those bandwidths by evaluating the FWHM of
the phase-matching function according to \cite{nistprogram},
extended to the case of PPLN. Here we consider the PDC
phase-matching function including the pump transverse distribution
\begin{equation}\label{phasematching}
\Phi(\omega_{\text{s}}, \theta_{\text{i}}, \theta_{\text{s}})=
\exp \left(- \frac{w_{\text{p}}^2 (\Delta k_{\text{x}}^2 + \Delta
k_{\text{y}}^{2})}{4}\right) \times  \left(\frac{\sin \Delta
k_{\text{z}} L}{\Delta k_{\text{z}} L}\right)^{2}.
\end{equation}
$\Delta k_{\text{x,y,z}}(\omega_{\text{s}}, \theta_{\text{i}},
\theta_{\text{s}})$ are the mismatch of the k-wavevectors $\Delta
k_{\text{x,y,z}}=(\mathbf{k}_{\text{p}}-\mathbf{k}_{\text{s}}-\mathbf{k}_{\text{i}})_{\text{x,y,z}}$,
in terms of the pump, signal, and idler k-vectors
\begin{eqnarray}\nonumber
\Delta k_{\text{z}}&=&\frac{ n(\omega_{\text{p}})
\omega_{\text{p}}}{c} -\frac{ n(\omega_{\text{s}})
\omega_{\text{s}}}{c} \cos\theta_{\text{s}}- \frac{
n(\omega_{\text{i}})
\omega_{\text{i}}}{c} \cos \theta_{\text{i}} - \frac{2 \pi} {\Lambda}\\
\nonumber \Delta k_{\text{x,y}}&=&\frac{ n(\omega_{\text{s}})
\omega_{\text{s}}}{c} \sin\theta_{\text{s}}+ \frac{
n(\omega_{\text{i}}) \omega_{\text{i}}}{c} \sin \theta_{\text{i}}.
\end{eqnarray}
$\Lambda$ is the poling period. To estimate the bandwidth selected
by the heralding arm fiber, we first find
$\theta_{\text{i}}^{\text{max}}$ that maximizes
Eq.(\ref{phasematching}), and then we consider for each wavelength
the function $f(\theta_{\text{s}}, \omega_{\text{s}})=1$ for
$\Phi(\omega_{\text{s}}, \theta_{\text{i}}^{\text{max}},
\theta_{\text{s}})> 0.5$ and $f(\theta_{\text{s}},
\omega_{\text{s}})=0$ for $\Phi(\omega_{\text{s}},
\theta_{\text{i}}^{\text{max}}, \theta_{\text{s}})< 0.5$. Finally
we invert the functions $f(\theta_{\text{s}} + \Delta
\theta_{1}/2, \omega_{\text{s}})$ and $f(\theta_{\text{s}}-\Delta
\theta_{1}/2, \omega_{\text{s}})$ to determine $\Delta_{1}$.

\begin{figure}[tbp]
\begin{center}
\includegraphics[width=15 cm]{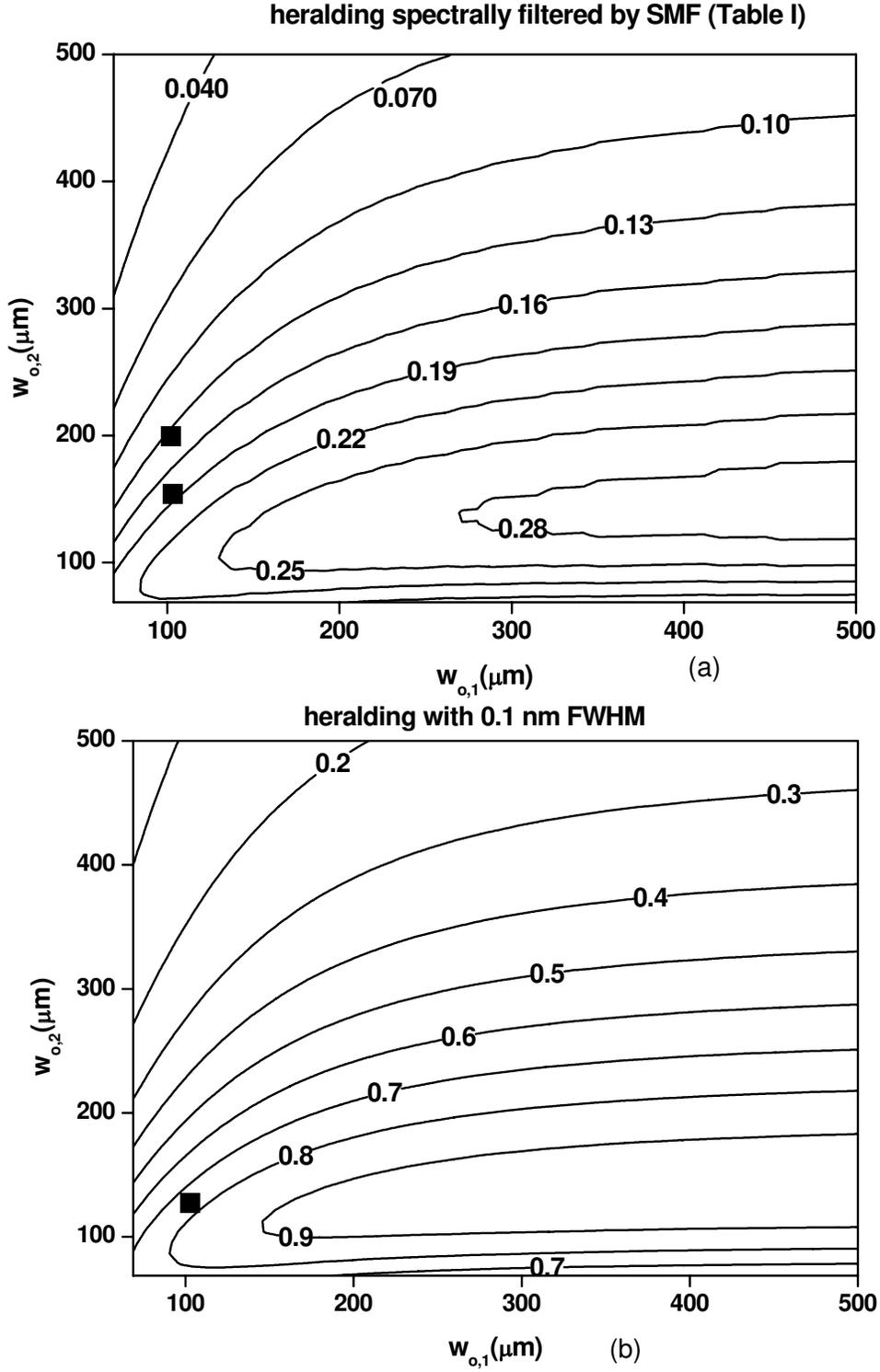}
\end{center}
\caption{Calculated $\chi_{\text{p}}$ versus heralding and heralded
waists (a) without an interference filter and (b) with spectral
filtering on the heralding arm of 0.1 nm FWHM. Estimated values in
the experimental conditions are given by the black squares.
$w_\mathrm{p}=144\mu m$ for both figures. }\label{figure2}
\end{figure}

We estimated the heralded arm bandwidth in the same way. This
estimation, which was found to be in agreement with the
experimental values, allowed us to extrapolate the bandwidths of
the heralded arm for other configurations. The discrepancy between
the estimated and measured values is likely due to using
Sellmeier's equations \cite{DHJ97} for undoped PPLN. Moreover, we
do not have precise knowledge of the crystal poling region length
(it maybe slightly shorter than full 5 mm crystal length), a
crucial engineering parameter. In Fig. \ref{figure2} we report the
estimated $\chi_{\text{p}}$ versus $w_{\text{o},1,2}$ for our
experiment's fixed pump waist, and crystal length, in two
spectral-selection configurations on the heralding arm, (a) the
SMF acts as a spectral filter with an average FWHM of about 2 nm
and (b) a FWHM 0.1 nm spectral filter (a monochromator was used
for this narrow bandwidth). The two experimental conditions for
the measurements in Table II are roughly indicated on Fig.
\ref{figure2}(a) by the squares, while the conditions for the
narrow band measurement is indicated on Fig. \ref{figure2}(b).
While the heralded arm is spectrally selected only by the fiber,
the large improvement achieved by narrowing the heralding
bandwidth indicates that spectral mode matching is more critical
than spatial mode matching (the only spatial requirement is that
$w_{\text{o},2}\cong w_{\text{p}}$, $w_{\text{o},1}>
w_{\text{p}})$. Using this we should be able to improve
$\chi_{\text{p}}$ by further narrowing the heralding arm
bandwidth.
\section{Experimental results}
Following the approach in section III.B, we measured the raw
detection efficiency in our setup at the best signal-to-noise
ratio of the heralding arm. Table II shows $\chi_{\text{D}}$
measured with $w_{\text{o},1}=$82 $\mu$m fixed ($d_{1}$=270 mm)
for 2 different positions of the heralded arm, corresponding to
$w_{\text{o},2}=$158 (197) $\mu$m ($d_{2}$=300 (380) mm). The
estimates and uncertainties calculated with this model and the
measured intrinsic statistical fluctuations show good agreement.
For each setup we tested the repeatability by applying our
alignment procedure \cite{MCD02} over several days. Fig. 3 shows
the measurement repeatability in agreement with the estimated
uncertainty at the 2 $\sigma_{\text{ML}}$ level. However, we point
out that the repeatability test was limited to collection lens
positions close to the source. We obtained theoretical values of
$\chi_{\text{p}}=20\%,$ and $13\%$, respectively for the waists in
Table II using only the SMF spectral selectivity. The decrease of
$\chi_{\text{p}}$ for bigger $w_{\text{o},2}$ is in agreement with
the measured decrease of $\chi_{\text{D}}$. Our theoretical model
for $\chi_{\text{p}}$ is very sensitive to the measurements of the
beam waists and the bandwidth estimates. Estimating the optical
transmittances at $\tau_{\text{opt}}$=65\% and
$\tau_{\text{SMF-lens}}$=83\% by an independent calibration and
using the measured $\chi_{\text{D}}$ in Table II, we were able to
consistently obtain $\chi_{\text{p}}=48 \% , 37$\%. On occasion we
measured $\chi_{\text{p}}$ much closer to one by inserting a
monochromator in the heralding arm, however the stability of our
setup did not allow repeatable results at this level.

At this stage, we note that the model can only be considered
qualitative due to the spectral approximation adopted. In our
approach we limited the calculation of the mismatch term to first
order in the transverse k-vector and the frequency, losing some of
the correlation between the frequency and the spatial variables in
the biphoton field \cite{CDS05}. To compensate, we introduced a
selection term due to the SMF spatial selection, but that works
best for the degenerate case where the PDC emission is broader and
this correction has less of an effect on the final mode matching
than in the nondegenerate case. That case requires a full
numerical solution.
\section{Conclusions}
We have shown the feasibility of using PPLN for single-photon
detector calibration at 1550 nm, as well as for a single-photon
single-mode source. We highlighted the need to maintain low dark
counts in the heralding arm to reduce statistical fluctuations. We
have shown that proper spatial mode matching and spectral mode
selection in the heralding arm is of paramount importance in
achieving the highest heralding efficiency.

This work was supported in part by ARDA, ARO, and DARPA/QUIST.





\newpage
\section{Figures and Tables}
\begin{table}[htb]\label{table20}
\centering\caption{Measured and estimated ${\Delta}_{1}$}
\begin{tabular}{cccc }
$d_{1}$ & $w_{\text{o},1}$ & ${\Delta}_{1} $ & $ {\Delta}_{1} $\\
  &   & measured & estimated\\
 (mm) &  ($\mu$m)  & (nm)& (nm)\\
\hline
 270 & 82 & 2.05  & 2.6\\
 400 & 105 & 1.87 & 2.3\\
 520 & 132 & 1.64 &2.1\\
  \\
\end{tabular}
\end{table}
\newpage
\begin{table}[htb]\label{table30} \centering\caption{Measured $\chi_{\text{D}}$ with the estimate,
maximum likelihood (ML) uncertainty and statistical fluctuations}
\begin{tabular}{ccc|ccc }
\multicolumn{3}{c|}{$w_{\text{o},2}=$158 $\mu$m} &
\multicolumn{3}{c}{$w_{\text{o},2}=$197 $\mu$m} \\
$\chi_{\text{D}} $ & $\sigma_{\text{ML}} $ & $\sigma_{\text{stat}} $
& $\chi_{\text{D}}$ &
$\sigma_{\text{ML}}$ & $\sigma_{\text{stat}} $\\
$(\%) $ & $(\%)$ & $(\%) $ & $(\%)$ & $(\%)$ & $(\%)$\\ \hline
2.508& 0.044 &0.040  & 2.015 & 0.030& 0.028 \\
2.555& 0.025 &0.019 & 2.056 & 0.028& 0.035 \\
2.584& 0.036 &0.031  & 2.107 & 0.034& 0.028 \\
2.489& 0.039 &0.037  & 1.996 & 0.032& 0.026 \\
\end{tabular}
\end{table}
\newpage
\newpage

\begin{figure}[htb]
\begin{center}
\includegraphics[angle=0,width=9 cm]{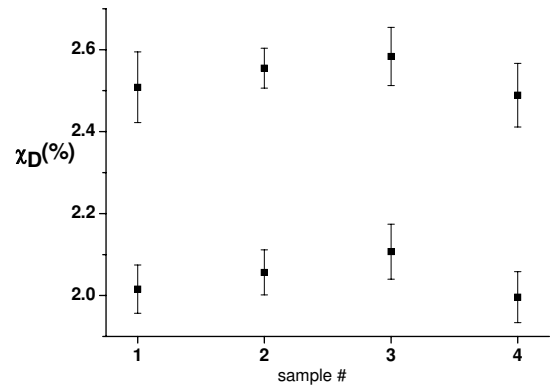}
\end{center}
\caption{Repeated measurements of $\chi_{\text{D}}$ for 2 setups
(\emph{k}=2 ML uncertainty shown). }\label{figure}
\end{figure}



\end{document}